# Low-Temperature Thermoelectric Performance and Optoelectronic Properties of Monolayer of WX$_2$N$_4$(X = Si, Ge).

Chayan Das[a], Dibyajyoti Saikia[a], Atanu Betal[a], Satyajit Sahu*

[a] Department of Physics, Indian Institute of Technology Jodhpur, Jodhpur 342037, India

## Abstract

Two-dimensional (2D) materials proved their suitability for thermoelectric applications due to specific quantum confinement and distinct density of states (DOS). New two-dimensional layered materials WX$_2$N$_4$(X = Si, Ge) are suitable for thermoelectric applications with a quite satisfactory value of ZT. Here we investigated the thermoelectric properties of the 2D monolayer of WX$_2$N$_4$(X = Si, Ge) by using Density Functional Theory (DFT) combined with Boltzmann Transport Equation (BTE). We obtained an outstanding thermoelectric figure of merit (*ZT*) of 0.91 at 400K for p-type WGe$_2$N$_4$, whether it showed a *ZT* value of 0.56 for n-type at the same temperature. On the other hand, the WSi$_2$N$_4$ showed significantly low ZT at room temperature. However, the *ZT* value increases significantly at higher temperatures. We examined the electrical property and discovered that the indirect bandgaps (BG) of WSi$_2$N$_4$ and WGe$_2$N$_4$ are 2.00 eV and 1.15 eV, respectively. In the deep ultraviolet (UV) and UV areas, they displayed very strong absorption. Due to their strong absorption, these materials may also be used in thermoelectric applications and (solar-blind) UV optoelectronic devices.

## 1. Introduction

The search for materials to fabricate highly efficient thermoelectric generators is trending research. Metal chalcogenides, alloys [1–6], and their oxides [7–9] made up the majority of the thermoelectric material regime. Bi$_2$Te$_3$ [10], SnSe [11], and PbTe [12] possess very high *ZT* value among them. After the discovery of graphene 2D transition metal dichalcogenides (TMDC) also attracted our attention because of their extraordinary optical [13], electronic [14], and thermal [14] properties. Recently new 2D monolayer of MoSi$_2$N$_4$ was deposited using the CVD method in centimeter scale [15]. So, we are motivated to explore the properties of these monolayers. These materials possess very high dynamical stability and very high mechanical properties [16].

Electricity can be generated from waste heat using these materials. Renewable energy sources are crucial in the current period since non-renewable energy sources are rapidly running out. In this work, we investigated the 2D monolayer of $WSi_2N_4$ and $WGe_2N_4$ using Boltzmann transport theory in association with DFT, and we found very high Seebeck coefficients ($S$). The seebeck coefficient of a material is the measure of the magnitude of induced voltage generated when a temperature difference is introduced between the two ends of the material. For efficient thermoelectric material, both the electrical conductivity ($\sigma$) and $S$ should be high, and the thermal conductivity ($k = k_{el} + k_{ph}$) should be low. The thermal conductivity $k$ has two components, $k_{el}$, and $k_{ph}$, which represents the contribution coming from electrons and phonons respectively. Generally, these 2D materials possess high mobility, excellent stability and good optoelectronic and piezoelectric properties [16]. These materials can be represented by N-X-N-M-N-X-N. Metal atom (M) was sandwiched in between two nitrogen (N) atoms, and the whole system was sandwiched between buckled honeycomb XN layer where X = Si or Ge. Here M is the transition metal atom (Mo, W, Ti, Cr, etc.). They can be synthesized using chemical vapor deposition (CVD). Yi-Lun Hong et al. synthesized $MoSi_2N_4$ using CVD [15]. Bohayra Mortazavi predicted mobility of 490 and 2190 $cm^2V^{-1}s^{-1}$ for p and n-type of monolayer $MoSi_2N_4$ [16]. Till now, many good researches have been carried out by researchers on 2D TMDC materials. Huang et al. and his group reported a very low $ZT$ value < 0.2, along with a high thermal conductivity of about 60 W/m K, with $MoSe_2$ monolayer. SD Guo and co-workers reported a $ZT$ value greater than 0.9 at 600K with $ZrS_2$ monolayer along with a thermal conductance of 47.8 W/K, respectively. Monolayer $MoS_2$ and $WS_2$ were also reported with very high $k_{ph}$ of 23.15 W/mK and 72 W/m.K [17]. But monolayer $HfS_2$ was reported with very low $k_{ph}$ of 2.83 W/mK along with a very good $ZT$ i.e. $ZT_{HfS_2} = 0.90$ [18]. Many researches are still going on to improve the $ZT$ factor of materials. In this work, we systematically investigated with great detail the electronic, optical, and thermoelectric properties of monolayers of $WSi_2N_4$ and $WGe_2N_4$ using DFT and BTE. In the $WGe_2N_4$ monolayer, we found an outstanding $ZT$ product of 0.91 for p-type (0.56 for n-type) at a very low temperature of 400K. Similarly, the $WSi_2N_4$ monolayer showed a poor $ZT$ of 0.26 for p-type (0.05 for n-type). Thus, we need a theoretical investigation of thermoelectric and optical properties to understand the physical and chemical properties that cause the huge difference in their efficiency towards thermoelectric application necessitates. $WSi_2N_4$ showed high absorption in the deep UV region (285 nm), whereas WGe$_2$N$_4$ showed strong absorption in the UV region (306 nm), making the materials usable in UV optoelectronic applications beyond the visible spectrum.

## 2. Methodology

We accomplished the first principle calculations using DFT with projector augmented wave (PAW) potentials [19,20] and Perdew-Burke-Ernzerhof (PBE) as generalized gradient approximation (GGA) [21] in Quantum Espresso (QE) package. To avoid the interaction of two layers, we kept a vacuum of 30 Å between two layers along the z-direction. The geometry optimization was performed using a 15×15×1 k-mesh grid. A wave function energy cutoff of 50 Ry and self-consistency of 10$^{-9}$ Ry were kept in all the calculations. The atoms were relaxed until the force convergent threshold of 3.8×10$^{-4}$ Ry was achieved. We used the Phonopy package combined with QE to evaluate the phonon dispersion band structure using 2×2×1 supercell with 9×9×1 k-mesh in the QE package. The optical properties were evaluated using the SIESTA package, which is implemented with Time-Dependent Density Functional Perturbation Theory (TD-DFPT) [22]. A 48×48×1 k-mesh was used for the calculation of the optical properties. We obtained the imaginary ($\varepsilon_i$) and real ($\varepsilon_r$) parts of dielectric functions using Momentum space formulation along with Kramers-Kronig transformation [23]. After, that we obtained the absorption coefficient (α), refractive index (η), and extinction coefficient (K) using the following equations.

$$\eta = \left[\frac{\left\{(\varepsilon_r^2+\varepsilon_i^2)^{1/2}+\varepsilon_r\right\}}{2}\right]^{1/2} \quad (1)$$

$$K = \left[\frac{\left\{(\varepsilon_r^2+\varepsilon_i^2)^{1/2}-\varepsilon_r\right\}}{2}\right]^{1/2} \quad (2)$$

$$\alpha = \frac{2K\omega}{C} \quad (3)$$

Here, $\varepsilon_r$, $\varepsilon_i$, $\omega$, and $C$ are real and imaginary parts of the dielectric function, frequency, and speed of light, respectively. Thermoelectric parameters were obtained using constant scattering time approximation from BoltzTraP code [24] using the Boltzmann transport equation.

$$\sigma_{l,m} = \frac{1}{\Omega}\int \sigma_{l,m}(\varepsilon)\left[-\frac{\Delta f_\mu(T,\varepsilon)}{\Delta \varepsilon}\right]d\varepsilon \quad (4)$$

$$k_{l,m}(T,\mu) = \frac{1}{e^2 T\Omega}\int \sigma_{l,m}(\varepsilon)(\varepsilon-\mu)^2\left[-\frac{\Delta f_\mu(T,\varepsilon)}{\Delta \varepsilon}\right]d\varepsilon \quad (5)$$

$$S_{l,m}(T,\mu) = \frac{(\sigma^{-1})_{n,l}}{eT\Omega} \int \sigma_{n,m}(\varepsilon)(\varepsilon - \mu)\left[-\frac{\Delta f_\mu(T,\varepsilon)}{\Delta \varepsilon}\right] d\varepsilon \qquad (6)$$

Using these equations, we obtained the transport properties. Here, $\sigma_{l,m}$, $k_{l,m}$, $S_{l,m}$, are the electrical conductivity, thermal conductivity, and Seebeck coefficient, respectively. Whereas $e, \mu, \Omega, T$ are the electron charge, chemical potential, unit cell volume, and temperature, respectively. The phono3py package combined with QE was used to evaluate the $k_{ph}$. In phono3py, a 2×2×1 supercell with 9×9×1 k-mesh was generated, and self-consistent calculations were performed using a default displacement of 0.06 Å.

## 3. Result and Discussions

### 3.1. Structural Properties and Stability

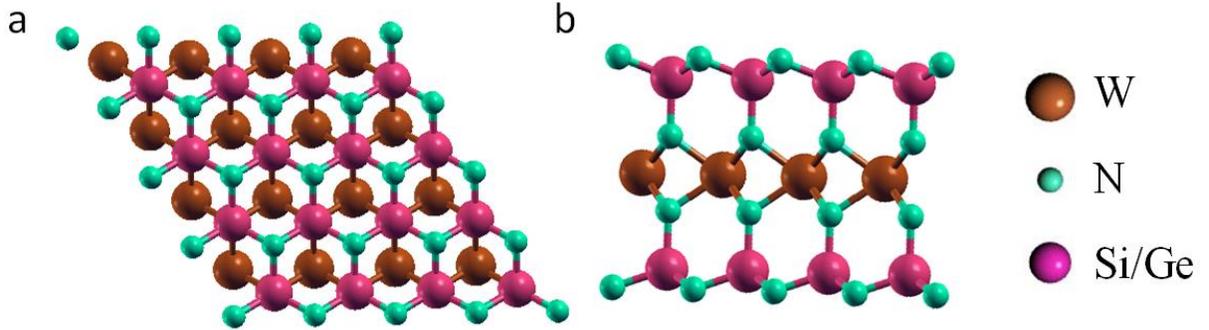

**Figure.1:** (a) Top and (b) side view of WX$_2$N$_4$ monolayer with honeycomb structure of 4×4×1 supercell.

WX$_2$N$_4$(X = Si, Ge) monolayer can be viewed as a WN$_2$ monolayer sandwiched between two honeycomb (SiN/GeN) layers. The three layers are stacked on each other. W atom is located at the center of a trigonal prism building block with six Si atoms, and the WN$_2$ layer is bonded to (SiN/GeN) layers via vertically aligned Si–N bonds (figure 1(a)). WX$_2$N$_4$ possesses a hexagonal primitive unit cell with space groups of P-6 m2 (No. 187) [16] (figure 1(b)). We relaxed the unit cells and obtained the lattice constant for WSi$_2$N$_4$ is a = b = 2.91 Å which matches exactly with previously reported results [16]. For WGe$_2$N$_4$, the parameters were a = b = 3.02 Å. The obtained lattice constants, bond lengths, and bond angles for both structures are shown in Table 1. Here we observed that the bond lengths were increased in the WGe$_2$N$_4$ compared to the bond lengths of WSi$_2$N$_4$ because the Ge atom has a larger atomic radius than the Si atom.

**Table-1:** Estimated lattice constants, bond lengths, thickness, and bond angles for monolayer of WSi$_2$N$_4$ and WGe$_2$N$_4$.

| Structure | a=b (Å) | d(W-N) (Å) | d(X-N) (Å) | Thickness(Å) | ϴ(W-N-W) | ϴ(N-X-N) | ϴ(W-N-X) |
|---|---|---|---|---|---|---|---|
| WSi$_2$N$_4$ | 2.91 | 2.10 | 1.75 | 10.94 | 72.96 ° | 106.62 ° | 126.47 ° |
| WGe$_2$N$_4$ | 3.02 | 2.12 | 1.85 | 11.54 | 69.65 ° | 109.50 ° | 124.83 ° |

The cohesive energy gives us information about the stability of the structure of the material. We evaluated the cohesive energy ($E_{ch}$) for both the structures given by the following formula: $E_{ch} = \{(E_W + 2 \times E_X + 4 \times E_N) - E_{WX_2N_4}\}/7$. Here $E_{WX_2N_4}$, $E_W$, $E_X$, and $E_N$ are the energy of the monolayer of WX$_2$N$_4$, the energy of a single W atom, the energy of a single X atom, and the energy of a single N atom, respectively. The cohesive energy per atom obtained for WSi$_2$N$_4$ and WGe$_2$N$_4$ monolayer was 0.97 eV, and 0.25 eV, respectively, which confirms that these structures are thermodynamically stable. To check the structural stability, we calculated the phonon dispersion curve (shown in figure 2) of WSi$_2$N$_4$ and WGe$_2$N$_4$, respectively, along the high symmetry path of Γ-M-K-Γ. No imaginary frequency was found in the phonon dispersion curve, which again confirms the thermodynamical stability of these structures. Since the unit cell of the monolayer contains seven atoms, there are in total twenty-one vibrational modes; the first three are acoustic modes, and the other eighteen the optical modes. The lower three branches correspond to the acoustic vibrational modes, and they are in-plane longitudinal acoustic (LA) mode, transverse acoustic (TA) mode, and out-of-plane mode (ZA). The upper eighteen modes are optical modes. It was observed that for the WGe$_2$N$_4$ monolayer, the optical bands overlapped with the acoustic modes, unlike the WSi$_2$N$_4$ monolayer, where the acoustic modes are independent.

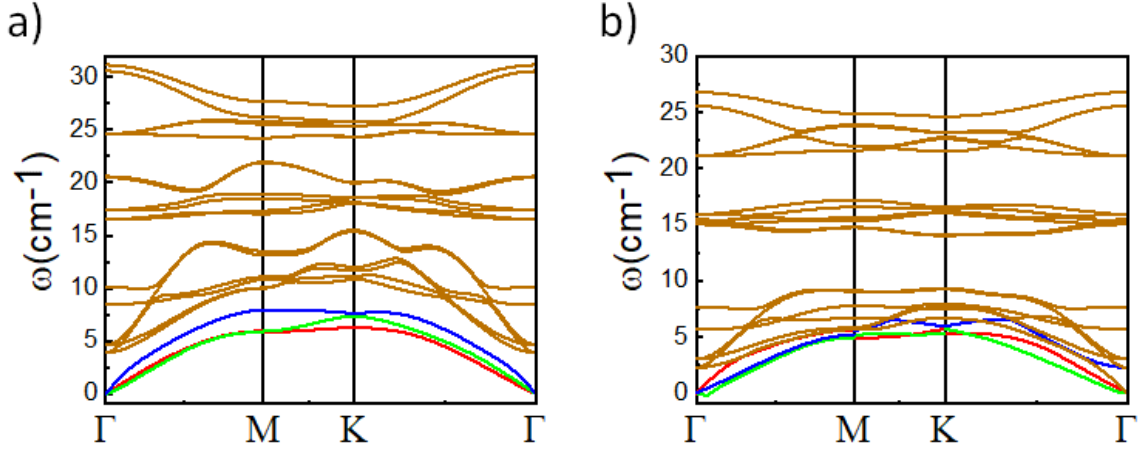

**Figure.2:** The phonon dispersion plot of the monolayer of a) $WSi_2N_4$ and b) $WGe_2N_4$. The red green and blue colored bands represent ZA, TA, LA bands of acoustic mode.

### 3.2. Electronic Properties

The electronic band structure of $WSi_2N_4$ and $WGe_2N_4$ monolayer was calculated along Γ-M-K-Γ path within an energy range from -4 eV to 4 eV. Figure 3a and 3b show the electronic band structure of $WSi_2N_4$ and $WGe_2N_4$, corresponding BGs of 2.00 eV and 1.15 eV, respectively, according to PBE. The obtained BG value for both materials matched precisely with the previous work by Bohayra Mortazavi et al. [16]. Both the band structures showed an indirect bandgap (BG), and for both the monolayer, the conduction band minima (CBM) are situated at K point. The valance band maxima (VBM) are located at Γ point for $WSi_2N_4$, but for $WGe_2N_4$, the VBM is situated near the Γ point, which was between the Γ and K points.

The DOS and Local density of states (LDOS) for different orbitals is shown in figure 4 for both monolayers. For $WSi_2N_4$, the contribution for VBM is mainly from the p orbitals of N and d orbitals of the W atom, but CBM is contributed primarily by d of the W atom, as shown in the in figure 4(a). For $WGe_2N_4$, the primary contribution for VBM is mainly from p of the N atom, and the contribution towards CBM is primarily from the d orbital of the W atom, as shown in figure 4(b). It is observed that for both the monolayer, the N contributes more to the valance band, and W contributes more to the conduction band. p orbital of Si and Ge contributes almost equally toward both the VBM and CBM.

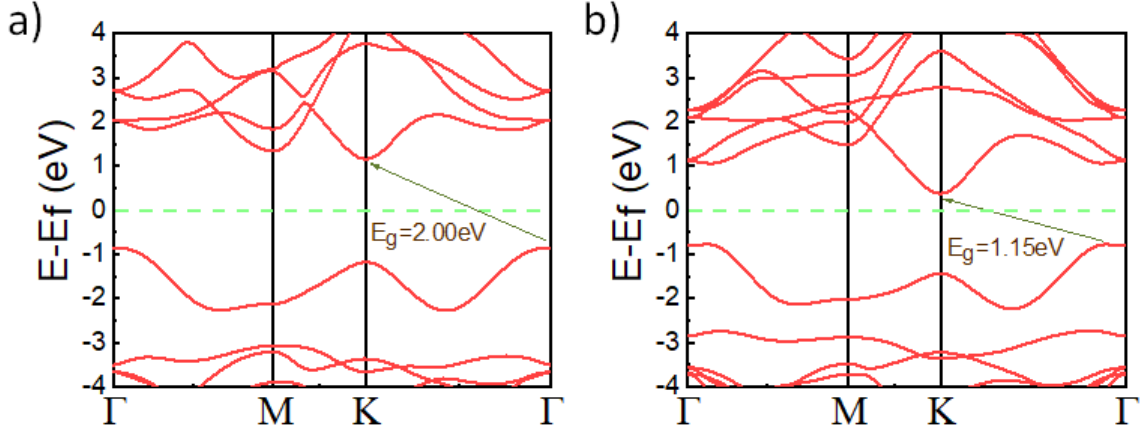

**Figure.3:** The Band structure of monolayer of a) WSi$_2$N$_4$ and b) WGe$_2$N$_4$, with PBE as GGA. The arrow signifies the BG and green dotted lines show the Fermi level.

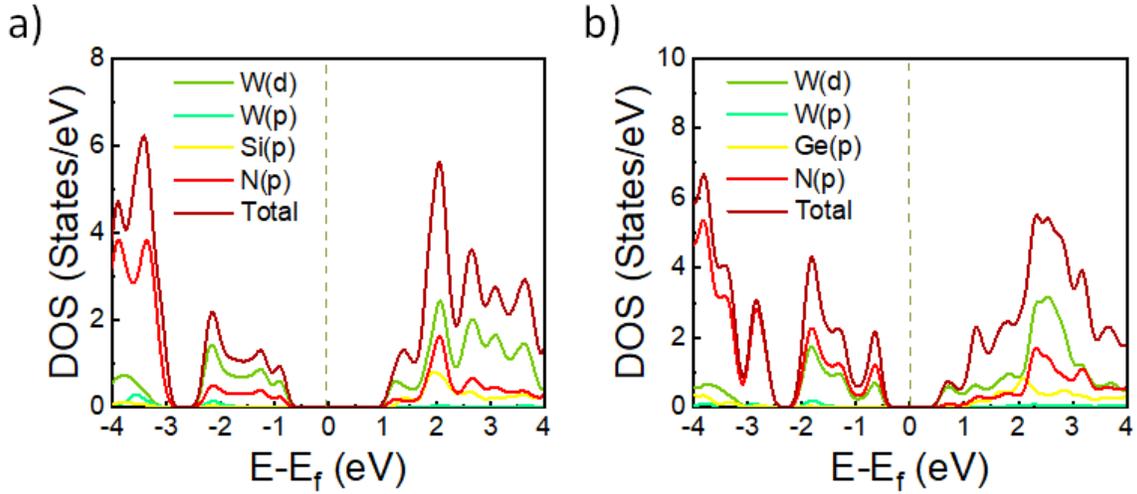

**Figure.4:** Total DOS with LDOS of the monolayer of a) WSi$_2$N$_4$ and b) WGe$_2$N$_4$ plotted as a function of energy.

### 3.3. Carrier mobility and relaxation time

The carrier mobility of electrons and holes for WSi$_2$N$_4$ and WGe$_2$N$_4$ was predicted by Bohayra Mortazavi et al. They predicted electron mobility ($\mu_e$) and hole mobility ($\mu_h$) was 320 cm$^2$V$^{-1}$s$^{-1}$ and 2026 cm$^2$V$^{-1}$s$^{-1}$ for WSi$_2$N$_4$ and 690 cm$^2$V$^{-1}$s$^{-1}$ and 2490 cm$^2$V$^{-1}$s$^{-1}$ for WGe$_2$N$_4$ monolayer. Effective mass ($m^* = \frac{d^2E}{dk^2}$) of specific charge carriers can be found from the band edge of band structures. The effective mass of the electron ($m_e$) can be determined by fitting a parabola at CBM and operating a double derivative. Similarly, the effective mass of an electron ($m_h$) can be determined by fitting a parabola at VBM and operating a double derivative. The relaxation time of the charge carrier was calculated by using the well-known formula $\tau = \frac{\mu m^*}{e}$. All calculated parameters for both crystals are listed in table 2.

| Crystal | $m_e(Kg)$ | $m_h(Kg)$ | $\tau_e \times (10^{-14})$ s | $\tau_h \times (10^{-14})$ s |
|---|---|---|---|---|
| WSi$_2$N$_4$ | $0.254 m_0$ | $0.617 m_0$ | 6.82 | 71.25 |
| WGe$_2$N$_4$ | $0.243 m_0$ | $0.512 m_0$ | 11.57 | 72.90 |

### 3.4. Optical properties

Optoelectronic device applications require an understanding of optical properties. The optical properties were investigated along the perpendicular direction of the plane. The $\varepsilon_r$ is obtained using Kramers-Kronig Transformation, and the $\varepsilon_i$ of dielectric function was calculated using momentum space formulation using proper matrix elements. The dielectric functions were plotted as a function of the energy of photons in figure 5(a) and (b). The $\varepsilon_i$ shows peak at 3.82 eV for WSi$_2$N$_4$, at 3.45 eV for WGe$_2$N$_4$. This can be explained by the band structure, as monolayer WGe$_2$N$_4$ has 0.85 eV lower BG compared to WSi$_2$N$_4$. The secondary peaks were found at 7.27 eV and 6.22 eV, respectively. The peaks were found at 2.55 eV and 1.95 eV for WSi$_2$N$_4$ and WGe$_2$N$_4$, respectively, for the $\varepsilon_r$. The absorption spectra are shown in figure 5(c), and the peaks were found at 4.35 eV and 4.05 eV with absorption coefficient ($\alpha$) of $2.39 \times 10^5$/cm and $2.23 \times 10^5$/cm for WSi$_2$N$_4$, and WGe$_2$N$_4$, respectively, which are of the order of SnI$_2$, SiI$_2$,[25] ZrS2 and ZrSSe [26]. Some secondary peaks are also observed near 8 eV for WSi$_2$N$_4$ and near 10 eV for WGe$_2$N$_4$ with an even higher absorption coefficient. The variation of refractive index ($\eta$) with energy is shown in figure 5(d). At zero energy, the refractive index is found to be 1.78, and 1.90 for WSi$_2$N$_4$ and WGe$_2$N$_4$, respectively. The secondary peaks are found at energy 2.70 eV and 2.10 eV for and at relatively high energy of 5.85 eV and 5.40 eV for WSi$_2$N$_4$ and WGe$_2$N$_4$. After 12 eV, the refractive indexes of WSi$_2$N$_4$ and WGe$_2$N$_4$ are almost the same.

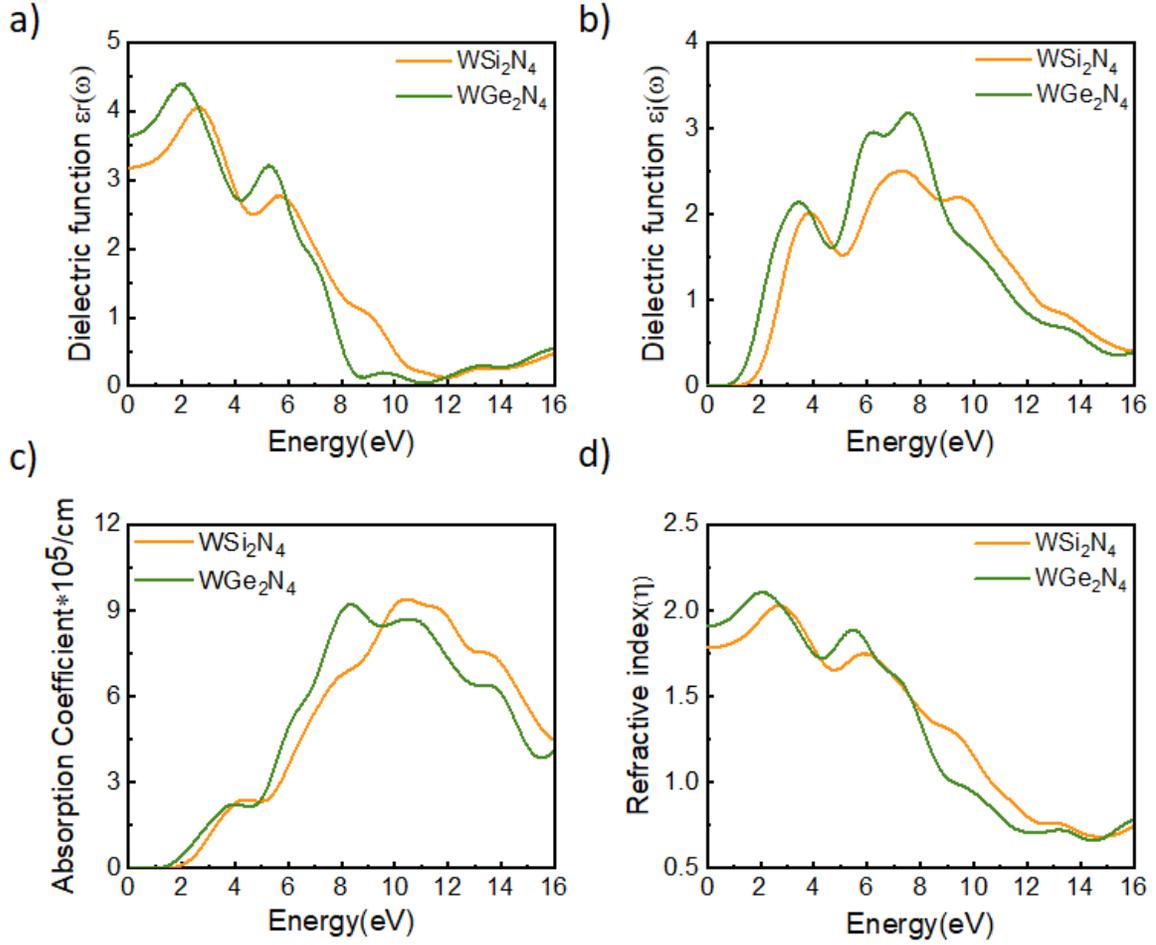

**Figure.5:** Representation of optical properties a)$\varepsilon_r$ , b) $\varepsilon_i$ , c) α, and d) η with respect to energy is shown.

### 3.5. Thermoelectric Properties

The plot of the Seebeck coefficient ($S$) with respect to µ at 300 K, 350 K, and 400 K for the monolayer of WSi$_2$N$_4$ is shown in figure 6a. The Seebeck coefficient ($S$) obtained for WSi$_2$N$_4$ monolayer at 350 K is 2705.29 µV/K for n-type carriers (µ > 0) and 2761.84 µV/K for p-type carriers (µ < 0). At 350 K, the $S$ with increased a little, but after that, it decreased at 400 K. The representation of relaxation time-scaled electrical conductivity ($\sigma/\tau$) as a function of µ is shown in figure 6b. No such change was found in ($\sigma/\tau$) for change in $T$ of 100K. The p-type WSi$_2$N$_4$ possesses much lower $\sigma/\tau$ compared to the n-type WSi$_2$N$_4$. The representation of the relaxation time-scaled power factor (PF = $S^2\sigma/\tau$ ) as a function of µ is shown in figure 6c for the WSi$_2$N$_4$ monolayer. The highest PF was obtained for WSi$_2$N$_4$ monolayer for n-type carriers (10.53×10$^{10}$

W/m$^2$Ks) at 400 K, and for p-type carriers, the highest PF was obtained (4.67×10$^{10}$ W/m$^2$Ks) at 400K. For the WGe$_2$N$_4$ monolayer, the variation of $S$, $\sigma/\tau$ and $S^2\sigma/\tau$ with respect to µ is shown in figure 6d, e, and f. Unlike the WSi$_2$N$_4$ monolayer, the $S$ gradually decreases with an increase in temperature. The highest power factor we obtained for the WGe$_2$N$_4$ monolayer is 8.60 ×10$^{10}$ W/m$^2$Ks, which is for the p-type carriers at 400 K, and for n-type carriers, the obtained power factor is 7.51 ×10$^{10}$ W/m$^2$Ks. So, the p-type doping is much more efficient in WGe$_2$N$_4$ towards thermoelectric application. The highest values of $S$, $(\sigma/\tau)$, $(S^2\sigma/\tau)$ for both the monolayers are listed in table 3.

**Table 3:** Calculated maximum a) $S$, b) $\sigma/\tau$, and c) $S^2\sigma/\tau$ for monolayer of WSi$_2$N$_4$ and WGe$_2$N$_4$.

| Crystal | Maximum Seebeck Coefficient(S) (µV/K) | | Maximum Conductivity ×10$^{19}$($\sigma/\tau$) (S/ms) | | Maximum Power Factor ×10$^{10}$ ($S^2\sigma/\tau$) W/m$^2$Ks | | Thermoelectric figure of merit (ZT) | |
|---|---|---|---|---|---|---|---|---|
| | p | n | p | n | p | n | p | n |
| WSi$_2$N$_4$ | 2761.84 | 2705.29 | 1.64 | 4.82 | 4.67 | 10.53 | 0.26 | 0.05 |
| WGe$_2$N$_4$ | 1961.78 | 1935.55 | 4.00 | 5.82 | 8.61 | 7.51 | 0.91 | 0.56 |

As the Seebeck coefficient ($S$) is directly proportional to BG[26], the $S$ for WSi$_2$N$_4$ monolayer was found to be higher compared to WGe$_2$N$_4$, which follows the BG trend for the two materials, i.e., WSi$_2$N$_4$ has the higher BG value compared to WGe$_2$N$_4$.

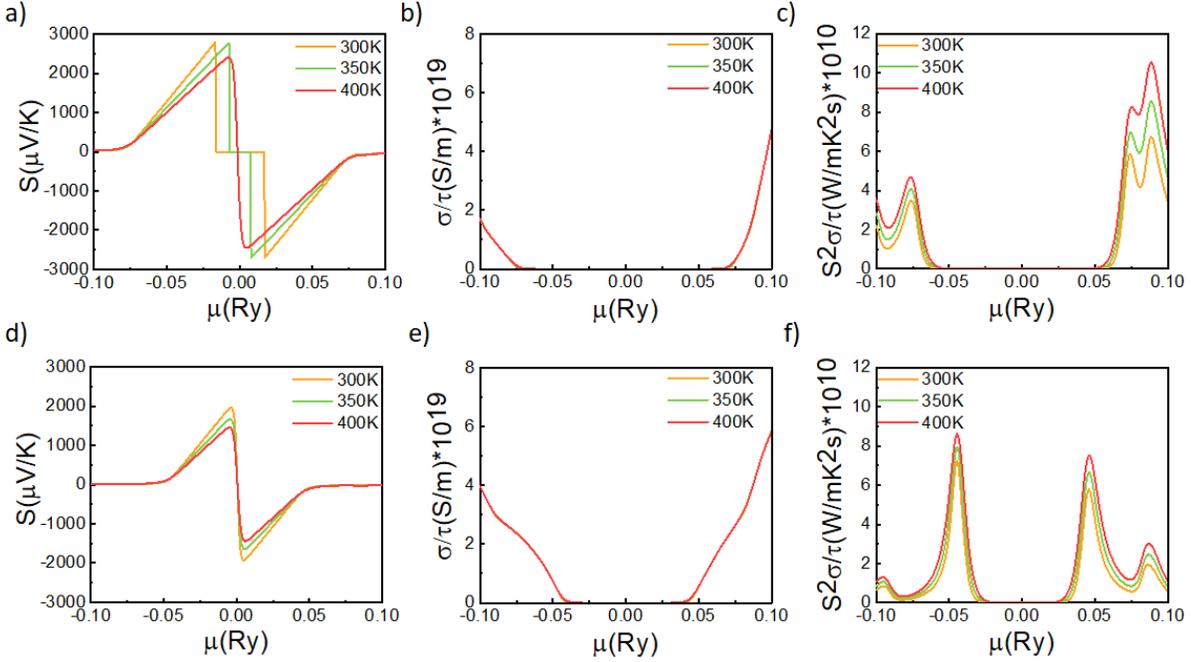

**Figure.6:** Representation of thermoelectric parameters as a function of µ for WSi$_2$N$_4$ monolayer a), b), and c), and for WGe$_2$N$_4$ monolayer, d), e), and f).

### 3.6. Lattice Thermal Conductivity ($\kappa_{ph}$)

The variation of $\kappa_{ph}$ with respect to $T$ for monolayer of WSi$_2$N$_4$, and WGe$_2$N$_4$ is shown in figure 7a, and 7b. Monolayer of WGe$_2$N$_4$ showed significantly lower $\kappa_{ph}$(3.63 W/mK) at 300 K, compared to WSi$_2$N$_4$ (41.9 W/mK) which are lower than the popular 2D TMDC like MoS$_2$ (34.5 W/(m.K)) [27], WS$_2$ (72 W/(m.K)) [17], SnS$_2$ (15.85 W/(m.K)) [14], SiSe$_2$ (15.85 W/(m.K)) [28] etc. A decrease in $\kappa_{ph}$ with an increase in $T$ is observed for both of the structures. The WGe$_2$N$_4$ monolayer possesses much lower thermal conductivity compared to monolayer of WSi$_2$N$_4$.

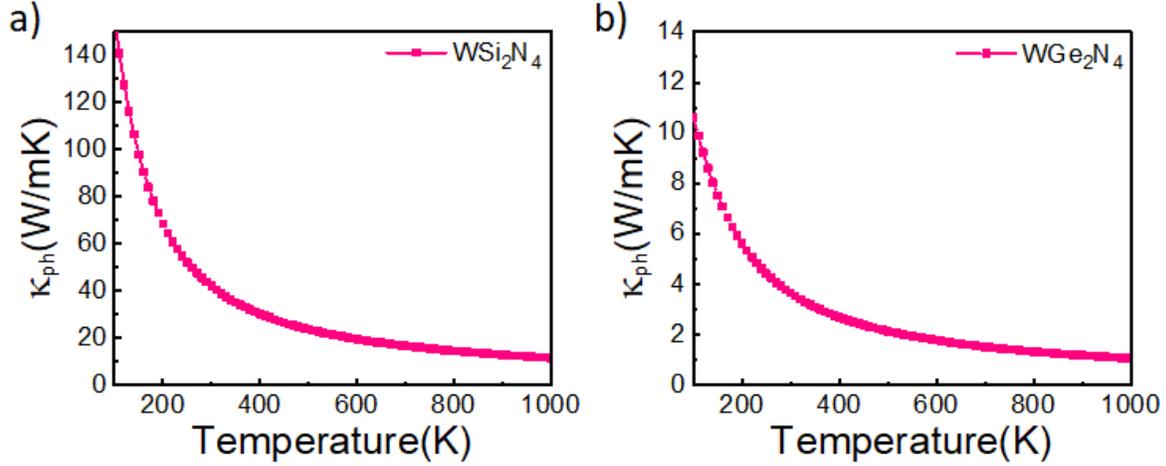

**Figure.7:** Plot of $\kappa_{ph}$ with respect to Temperature(K) of monolayer $WSi_2N_4$ and $WGe_2N_4$.

The change of phonon lifetime ($\tau$) and group velocity ($G_v$) as a function of frequency for both the monolayers is shown in figure 8. The maximum phonon lifetime of $WSi_2N_4$ (figure 8a) is observed to be much higher than the maximum phonon lifetime of the $WGe_2N_4$ monolayer (figure 8b). Which is low compared to $MoS_2$ and $WS_2$ [29,30]. The maximum phonon group velocity obtained for the $WSi_2N_4$ monolayer is about 11.1 Km/s (figure 8c) which is due to the out-of-plane acoustic mode. Similarly, the maximum obtained $G_v$ for $WGe_2N_4$ is about 7.5 Km/s (figure 8d) due to the out-of-plane acoustic mode combined with optical modes. As the $\kappa_{ph}$ is proportional to $G_v$, and $\tau$, the change in $\kappa_{ph}$ of the monolayers can be explained easily. As both the $\tau$ and $G_v$ for $WSi_2N_4$ monolayer is much higher than the $WGe_2N_4$, the $\kappa_{ph}$ is much higher in $WSi_2N_4$. As from the phonon band structure we can observe that for $WSi_2N_4$ the acoustic and optical phonon bands are not overlapped, which leads to the increment in $\kappa_{ph}$ for $WSi_2N_4$ monolayer. On the other hand, for the $WGe_2N_4$, the acoustic and optical phonon bands are overlapped which leads to the decrement of $\kappa_{ph}$ for $WGe_2N_4$ monolayer.

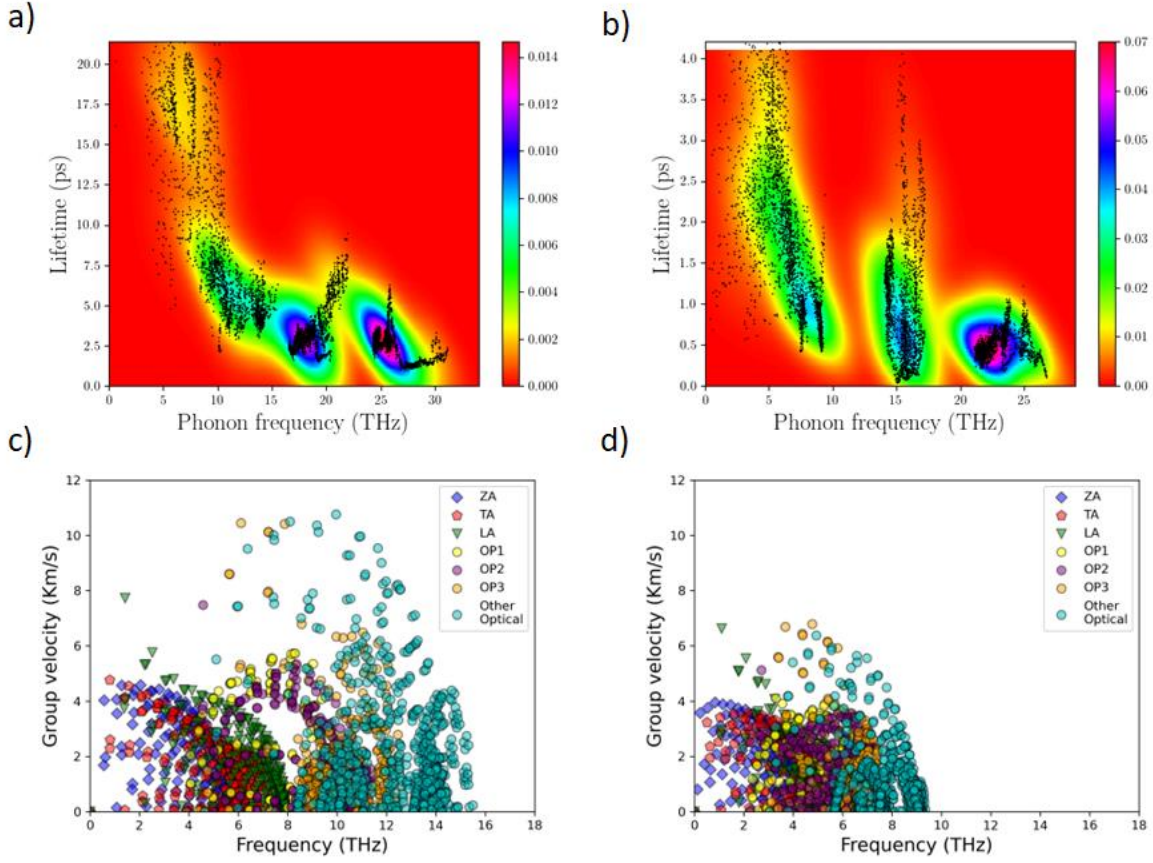

**Figure.8:** Variation of phonon lifetime with phonon frequency (a) for $WSi_2N_4$ and (b) $WGe_2N_4$ monolayer. Variation of group velocity ($G_v$) for various acoustic and optical modes with the frequency for (c) $WSi_2N_4$ and (d) $WGe_2N_4$ monolayer.

### 3.7. Thermoelectric figure of merit (*ZT*)

The parameter *ZT* signifies the efficiency towards the thermoelectric effect along with the quality of a material. To get a high *ZT* value, we need a high value of $\sigma$ and a much low value of $k$ ($k_{el} + k_{ph}$) as represented in equation 7. The Thermoelectric figure of merit (*ZT*) is defined as

$$ZT = \frac{S^2 \sigma T}{\kappa_{el} + \kappa_{ph}} \qquad (7)$$

Where $S, T, \sigma, k,$ and $k_{ph}$ are the same as defined earlier. The obtained *ZT* product is represented as a function of μ for the monolayer of $WSi_2N_4$ and $WGe_2N_4$ at 300 K, 350 K, and 400 K, as shown in figure 9(a) and (b), respectively. Among them, $WGe_2N_4$ monolayer shows the highest *ZT* value, i.e., 0.91 for p-type at 400 K and 0.56 for n-type, respectively. On the other hand, the $WSi_2N_4$ monolayer showed a very low *ZT* value of 0.26 for p-type and 0.05 for n-type at 400 K. However, at higher temperatures like 900K, the *ZT* value reaches around 0.69 (supplementary) for the p-type of $WSi_2N_4$ monolayer. Both the structures showed higher *ZT* values towards p-type doping compared to n-type doping. So, for both the structures, the p-type doping (compared to n-type doping) showed a higher *ZT* value which signifies the effectiveness of p-type doping compared to n-type. The *S* of $WSi_2N_4$ is higher rather than $WGe_2N_4$, but the $\sigma$ of $WGe_2N_4$ is higher than $WSi_2N_4$, so the power factor remains comparable, but the $k_{ph}$ played the crucial role in making the $WGe_2N_4$ monolayer an excellent material towards thermoelectric effect.

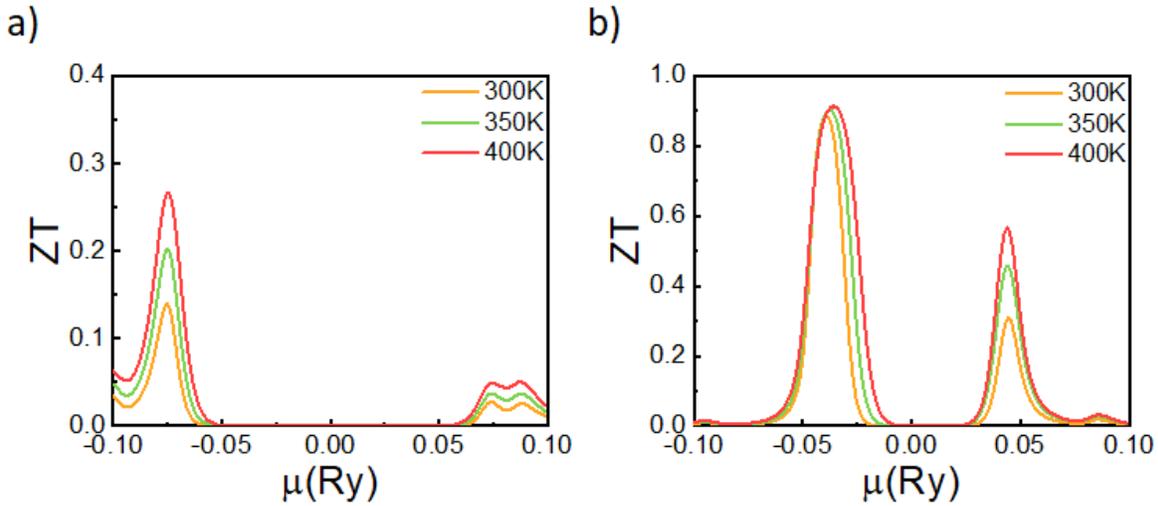

**Figure.9:** Plot of *ZT* at different temperatures (*T*) with respect to chemical potential (µ) for the monolayer of (a) $WSi_2N_4$ and (b) $WGe_2N_4$.

## 4. Conclusion

We calculated the electronic, optical, and thermoelectric properties using DFT and BTE. The absence of imaginary frequency in phonon dispersion curves proved the dynamical stability of both structures. The *ZT* and $k_{ph}$ for $WGe_2N_4$ are found to be outstanding for low-temperature operation. And $WSi_2N_4$ is not much efficient in the thermoelectric application. An excellent *ZT* product found is 0.91 for the p-type of $WGe_2N_4$ at 400 K. So, the $WGe_2N_4$ monolayer with p-type doping can significantly increase the thermoelectric performance. So, $WGe_2N_4$ can be used for next-generation low-temperature excellent thermoelectric devices to generate electricity from waste heat.

# Acknowledgement:

We are thankful to the Department of Science and Technology (DST), India for supporting us through the INSPIRE program and the Ministry of Human Resource and Development (MHRD). We are also grateful to the Indian Institute of technology Jodhpur for providing the infrastructure to carry out the research.